# A heuristic view about the evolution and species

Yan Huang

Department of Horticulture, Zhejiang University,
Huajiachi Campus, 310029 Hangzhou, People's Republic of China

**Abstract**

The controversy concerning both the definition of the species and methods for inferring the boundaries and numbers of species has occupied biologists for centuries, and the debate itself has become known as the species problem. The modern theory of evolution depends on a fundamental redefinition of "species". Here we show that based on the model of evolutionary continuum combined with fuzzy theory that the evolution system is a uncountable infinite set and species is a fuzzy set, the contradiction between discrete biological entities and continuous evolution system is solved, i.e. when a species evolved, the individuals scattered in space but continuously distributed on time sequences. Moreover, the calculation methods for species are suggested both in theory and practice.

## 1   Introduction

Traditionally, a species was seen as a group of organisms exemplified by a "type specimen" that bears all the traits common to this species[1]. However, biologists became conscious of species as statistical phenomenon[2-4] and not as categories which is counterintuitive since the classical idea of species is still widely-held. It is known that Darwin's evolutionism – as usually understood at the present time—when applied to species, leads to indefinable concept which is not valid generally in systematics. Since the continuing evolution of species, Darwin[2] treated it as the concept of pure mode with   differential degree. In any case the theory requires a correction, and I shall attempt in the following to accomplish this on the basis of the theory of fuzzy set and quantum which I developed. For this purpose it will be necessary first to find in set of conditions leading to Darwin's evolutionism that term which can be changed; thereafter it will be a matter of removing this term from set and making an appropriate substitution for it.

   Dobzhansky[3] and Mayr[4,5] emphasized 2 themes in the study of animal systematics:  ① Diversity, namely phenotypic diversity which was the premier and the fundamental subject of evolutionary biology.  ② Discontinuity, namely breaking in smoothing continuous area of phenotypic diversity, and this continuous area was the accumulation results for adaption through natural selection. Few biologists have noted that Darwin set up the argument in his book that the obvious discontinuity presented between closely interrelated contemporary species. Many biosystematist became conscious of the importance of discontinuity in geographical distribution. And this distribution discontinuity and phenotypic reproduction discontinuity of species in pairs or in groups closely related but with obvious differentiation interdependence influenced.

   In fact, it seems to me that the contemplation on "indefinable species", discrete biological entities and continuous evolution system can be better understood on the model of evolutionary continuum combined with fuzzy theory that the evolution system is a uncountable infinite set and species is a fuzzy set. According to the assumption considered here, when a species evolved, the individuals scattered in space but continuously distributed on time sequences.

In the following, I shall communicate the thinking and the the facts which led me to this conclusion, in the hope that point of view to be given may turn out to be useful for some researchers in their investigations.

## 2 Model of Evolutionary Continuum

In the basic set theory[6], the discussion about infinity given us the hint of mathematical model of biological evolution which take the whole evolution as a continuum. Continuum formulation is very simple. If a system of infinite elements can be arranged out one by one, for example, natural number system, it can be arranged from small to large: 1, 2, 3 ..., then this infinite system can be called countable. However, there is an uncountable system, the elements can not be listed, such as real number system, randomly give a real number - a, the successor number could not be found. Because if b is the successor number to a, we can always find a number, such as (a + b) / 2, it is between a and b, and this process is endless, that is, infinite numbers are between two arbitrary real numbers. This uncountable infinite system is the so-called mathematical continuum.

The dominant scientific theory of 20th century was quantum mechanics which was on the basis of quantisation, and the universe is recognized with regard to a countable infinite system in fact[7]. The biological evolution system is discrete or continuous, and scientific laws is probabilistic or causal, these two propositions are equivalent in nature. Therefore, the establishment of a continuum evolution model for the scientific basis theory is to establish a unified scientific theory of causality. How to develop the discrete entities to a scientific theory of continuity? The key is to convert the countable infinite system to the continuum. And this problem has ready answer in set theory. Set theory indicates that any power set is greater than the original set, and the power set of countable infinite set is the continuum. It can be said that this is a mystery of the universe, it has been bred in set theory, however, has not been able to arouse people's attention.

In fact, discrete biological entities and the continuous evolution system are not contradictory. In accordance with the set theory, a countable infinite set with infinite combinations creats power set, and this power set is a continuum. Then can the various discrete subsystems i.e. biological entities in evolution system infinite combinate? With the idea of probability theory, it is obviously impossible, but as evolution is a function of time which make this combination possible. In other words, if discreteness is "linear", that is, the entities in evolution is only superposed entities, then the evolution system is countable, but if the evolution investigated on time sequences, then the evolution system is continuous. In this way, the complete evolution continuum will be demonstrated in front of people.

Assume X as the set of biological entities(countable infinite set), and all the ordinary subsets of X construct a set recorded as $P(x)$ i.e. the evolution system.

Evolutionary continuum possess three main characteristics, first - the unity, that is, all existences in evolution can be transformed into each other and share with the same nature; second - the continuity, that is, all existence in evolution can be quantized, and to form the continuum; third - completeness, that is the local development of evolution converge to the overall evolution.

## 3 A Priori Species on the Fuzzy Theory[8]

The following discussion is on X - the set of biological entities(countable infinite set), namely the universe, S - the finite interval on X, and the set generated by all the ordinary subsets of X is denoted as $P(x)$ - the evolution system.

Because the concept of species is the product in the process of people's subjective consciousness to cognize objective things. To determine the membership of biological entities to the fuzzy set Ã of species, it is inevitably to reflect the judgment and reliability of person's subjective consciousness to the objective things, which is subjective. At the same time, species concept is the reflection of objective things in the minds of people, it is subject to objective constraints and restrictions, from that point of view it is objective. The substance of discussion above is the contradiction property of species. Herein, we may define the fuzzy set Ã of species as follows:

Set Ã: S → [0,1], s → Ã (s)

It is called the fuzzy set Ã on S, while the function Ã (•) as the membership function of fuzzy set Ã. Ã (s) is as yet known membership of s to fuzzy set Ã. Then Ã can be expressed as:

Ã = ((s, Ã (s)) | s ∈ X)

Now according to S as a countable finite set Ã is given by:

$$\tilde{A} = \sum_s \frac{\tilde{A}(s)}{s}$$

The definition of species in fuzzy set is therefore Ã; and this gives us the meaning of the concept.

## 4    Closest Principles to Judge Species[9]

Assume species $\tilde{A}_i$ ∈ F (x) (i = 1,2, ···, n), and each species is described by m characters, respectively $x_1, x_2, \cdots, x_n$, so there is n × m fuzzy sets $\tilde{A}_{ij}$ ∈ F($x_j$) (i = 1, 2, ···, n; j = 1, 2, ···, m) to represent different characteristics of species. On the other hand, assume set $\tilde{B}$ ∈ F(U) to the object to identify, of which m fuzzy character sets are $\tilde{B}_j$ ∈ F($X_j$) (j = 1, 2, ···, m). For ∀ i ∈ (1, 2, ···, n), we obtaine

$$S_i = \min\{D(\tilde{B}_1, \tilde{A}_{i1}), D(\tilde{B}_2, \tilde{A}_{i2}), \cdots, D(\tilde{B}_m, \tilde{A}_{im})\}$$

where $D(\tilde{B}_m, \tilde{A}_{im})$ is lattice degree of approaching to $\tilde{A}_{im}, \tilde{B}_m$.

If $S_{i0}$ = max{$S_1, S_2, \cdots, S_n$}

Then the object to identify is considered to be closest to category $i_0$, that is, shall be bolong to the species $i_0$.

## 5    A Posteriori Species in Multiphase System

The aim of orthodox systematics is to yield taxonomic groups such as species which bring together organisms with the highest proportion of similar attributes[10]. This is the logical basis of what I somewhat loosely call 'natural' species, for it enables me to make the greatest number of generalization about the organisms which are grouped into one species. I hold the view that a 'natural' or orthodox species is general arrangement intended for general use by all kinds of scientists.

Let us take a system of co-ordinates in which one dimension for each character. In order to render our presentation more precise and to distinguish this system of coordinates verbally from others, we call it the `` multiphase system" which is a ``stationary system." If an organism occupies one point in this multiphase system of co-ordinates, its position can be defined relatively thereto by the employment of rigid standards of measurement and the methods of Euclidean geometry, and can be expressed in Cartesian co-ordinates.

Organism relationships are to be evaluated purely on the basis of the resemblance existing in the materrial at hand. As they take into account the mode of origin of the observed resemblances or the rate at which such resemblances have increased or decreased in the past, the relationships are therefore dynamic on time sequences. As the more characters we study the more information we will accumulate. However, as we include more and more characters the rate of gain of new information for purposes will decrease. The resemblance would become more stable as the number of characters increases, and would eventually approach that parametric proportion of matches which we would obtain if we were able to include all the characters.

Characters should be unit charcaters, or if they are multiple they should be broken down into unit characters. It is possible to interpret and define these in terms of bioinformation theory. The organizational level of a unit character can differ considerablely from one character can differ considerably from one character to another and with advances in our knowledge of them.

For the computation of resemblance the Euclidean distance between the entities can be calculated by a simple formula. The Euclidean distance between points $P = (p_1, p_2, \ldots, p_n)$ and $Q = (q_1, q_2, \ldots, q_n)$, in multiphase system, is defined as:

$$\sqrt{(p_1^2 - q_1^2)^2 + (p_2^2 - q_2^2)^2 + \cdots + (p_n^2 - q_n^2)^2} = \sqrt{\sum_{i=1}^{n}(p_i - q_i)^2}$$

If we give the values of organisms' co-ordinates as functions of the time, it is proper to study species in multiphase system. As all biological entities in multiphase system forming a hypersurface, the concept of species should be defined as the stagnation point on hypersurface and its neighborhood. Combined theory of multivariate function and topology, the calculation method for species is as follows:

a): construct a function;

All points marked in multiphase system fit a hypersurface, or by means of dimension reduction construct a surface or curve.

b): calculate stagnation points;

Stagnation point p is of the points where the first derivative of function is zero (i.e. zero gradient).

c): calculate neighborhood;

p is a point on X, the neighborhood of p is set V, it contains the open set U containing p,

p ∈ U ⊆ V.

If S is a subset of X, the neighborhood of S is set V, it contains U containing the open set S that. Thus set V is the neighborhood of S, if and only if it is the neighborhood of all points on S.

**References**

[1] De Queiroz K (2007) Species concepts and species delimitation. Syst Biol. 56 (6):879-886.
[2] Darwin C (2004) The Origin of Species. CRW Publishing Limited.
[3] Dobzhansky T (1st ed., 1937; 2nd ed., 1941) Genetics and the Origin of Species. Columbia University Press.
[4] Mayr E (1942) Systematics and the Origin of Species, from the Viewpoint of a Zoologist.

Harvard University Press.

[5] De Queiroz K (2005) Ernst Mayr and the modern concept of species. Proc Natl Acad Sci U.S.A. 102 Suppl 1: 6600-6607.

[6] Hrbacek K, Jech TJ (1999) Introduction to Set Theory. CRC Press

[7] Griffiths DJ (2006) Introduction to Quantum Mechanics. Pearson Prentice Hall.

[8] Zimmermann HJ (2001) Fuzzy Set Theory--and Its Applications. Springer.

[9] Pal SK, Pal A (2001) Pattern Recognition: From Classical to Modern Approaches. World Scientific.

[10] Sneath PHA, Sokal RR (1962) Numerical taxonomy. Nature 193: 855-860.